\documentclass[twocolumn,showpacs,preprintnumbers,
amsmath,amssymb,aps,prd,nofootinbib,superscriptaddress,
eqsecnum]{revtex4}
\usepackage{graphicx,color}
\makeatletter
\renewcommand{\p@subsection}{}
\makeatother

\newcommand{\Slash}[1]{\ooalign{\hfil/\hfil\crcr$#1$}}

\begin{document}

\title{
Transport coefficients near  chiral phase transition }

\author{Chihiro Sasaki}
\affiliation{%
Technische Universit\"at M\"unchen,  D-85748 Garching, Germany}
\author{Krzysztof  Redlich}
\affiliation{%
Institute of Theoretical Physics, University of Wroclaw, PL--50204 Wroc\l aw, Poland}
\affiliation{%
Institute f\"ur Kernphysik, Technische Universit\"at Darmstadt, D-64289 Darmstadt, Germany}

\date{\today}

\begin{abstract}
We analyze   the transport properties of relativistic fluid composed of constituent quarks at
finite temperature and density. We focus on   the shear and bulk viscosities and study their
behavior near  chiral phase transition. We model the constituent quark interactions through the
Nambu--Jona Lasinio  Lagrangian. The transport coefficients are calculated within kinetic theory
under relaxation time approximation including in--medium modification of quasi--particles
dispersion relations. We quantify  the influence of the order of chiral phase transition and  the
critical end point on dissipative phenomena in such a medium. Considering the changes of shear and
bulk viscosities along the phase  boundary  we discuss their sensitivity to  probe the existence of the  critical end point.

\end{abstract}


\setcounter{footnote}{0} \maketitle

\section{Introduction}
\label{sec:int}

The shear $\eta$ and bulk $\zeta$ viscosities are parameters that quantify  dissipative processes
in the hydrodynamic  evolution of  fluid. Furthermore, it has been
argued that $\zeta$ and $\eta$ are    sensitive to  phase transitions 
in a medium~\cite{hydro,kapusta,eta:njl,hadronic,mclerran,rhic,chiral,tuchin,polyakov,karsch}. Indeed for certain materials, e.g.
helium, nitrogen or water, the shear viscosity to entropy ratio $\eta/s$ is known experimentally to
exhibit  a minimum at the phase transition \cite{mclerran}.
On the other hand, the bulk viscosity $\zeta/s$
was argued to be large   at the critical  point
\cite{chiral,tuchin,polyakov,karsch}.

In QCD of particular interest are properties of transport coefficients near the critical line where deconfinement and chiral phase transition sets in. This is because any change of the bulk and shear viscosities near $T_c$  modifies    the hydrodynamic evolution of the QCD medium and influences phenomenological observables  that characterize its expansion dynamics. 

The recent Lattice Gauge Theory (LGT)  calculations in pure SU(3)
seem to be consistent with the expectation of
decreasing $\eta/s$  and increasing $\zeta/s$ toward the first order  deconfinement  phase transition approaching from the 
high temperature phase~\cite{lgts,lgtb}. The properties of  the energy-momentum tensor in the vicinity
of the second order  phase transition  and its importance
for  the analysis
of transport coefficients and their critical behavior have been also calculated and discussed
within LGT 
for  (3+1)-dimensional SU(2) gauge theory \cite{lgtk}.
All these results show that  the  transport coefficients are of particular interest to quantify the
properties of strongly interacting relativistic fluid  and its phase transition \cite{mclerran}.

In   the theory of critical phenomena, describing
dynamical processes near  critical point and their universal behavior, 
 the transport coefficients in  the second order phase transition can be quantified by a dynamical critical exponents $z$ \cite{hh,onuki}. These exponents are   common for all models belonging to the same dynamical  universality class.  In general, the critical exponent $z$ determines  the critical slowing down of the system
relaxation time  $\tau\sim \xi^z\sim |T-T_c|^{-\nu z}$  near the critical temperature $T_c$  with
$\xi$ being correlation length and $\nu$ its static critical exponent. It was argued that within  the   classification given by  Halperin  and Hohenberg
 the  QCD  critical end point  (CEP) belongs to the universality class of the model H~\cite{ss}.  Consequently, both shear and bulk viscosities  are expected to  diverge at the CEP as $\eta\simeq \xi^{z_\eta}$  and  $\zeta\simeq \xi^{z_\zeta}$  with dynamical critical exponent $z_\eta\simeq 1/19$ and $z_\zeta\simeq 3$ respectively \cite{onuki}.

The  analysis of experimental data obtained in heavy ion collisions at RHIC showed  that
the evolution of the quark gluon plasma (QGP) is well described by  
nearly ideal hydrodynamics \cite{hadro}. It has  been even argued  that 
 the QGP is in fact  the most nearly ideal fluid known, with 
shear  viscosity close to the conjectured lower bound in any system \cite{ds,DG}. Clearly, in the evolution of QCD medium its transport properties are changing when approaching the phase boundary. Thus, it is  of importance    to   quantify  the change of  shear and bulk viscosities with thermal parameters  near the phase transition expected in strongly interacting medium.

In this paper we 
explore   the transport properties of relativistic fluid  composed of constituent quarks at
finite temperature and density. The dynamics  of medium constituents is described  by the chirally invariant four--quark interactions within Nambu--Jona-Lasinio (NJL) model \cite{nambu,review}.
The model correctly describes the universal critical  behavior of physical observables  with static critical exponents that belongs to the QCD universality class with respect to chiral symmetry.

 To calculate  the change  of the shear and bulk viscosities with thermal parameters,   we assume that  the system appears  near equilibrium and we apply the kinetic theory under relaxation time approximation \cite{HK,gavin,sr,arnold,moore}. The relaxation time of quarks is obtained from the
thermally averaged total cross-section of elastic scattering in the dilute gas approximation \cite{eta:njl}.
 
  In the NJL model under  mean field dynamics the interactions of quarks lead to the quasi--particle description of thermodynamics with their masses $M(T,\mu)$ which  are temperature $T$- and density $\mu$-dependent \cite{review,sfr:prd}.  The dynamically generated mass acts as  an order parameter for chiral phase transition.  In our  calculations of transport coefficients, following the  method described in Ref. \cite{sr}, the constituent quark mass  $M(T,\mu)$ is consistently built in.
We quantify  the influence of the order of the chiral phase transition and  the
critical end point   on the shear and bulk viscosities. 
We show  that  the bulk viscosity is strongly increasing whereas the shear viscosity is decreasing   when approaching the CEP from the side  of chirally symmetric phase.  These results are consistent with the expectation that the bulk viscosity could dominate  dissipative hydrodynamics near CEP \cite{chiral,tuchin}.   Considering the changes of $\eta$ and $\zeta$  along the phase  boundary  we discuss their sensitivity to  probe  the existence of the  critical end point.

Within the kinetic theory and under linear response  and relaxation time approximation used in this calculation, there is no   access to the critical dynamics near second order transition where the properties of shear and bulk viscosities are quantified by dynamical critical exponents. In addition, in this formulation,  the transport coefficients are finite with respect to static critical exponents \cite{sr}. Thus, the behavior  of  $\eta$ and $\zeta$  is entirely governed by the non-singular part of the partition function. Consequently,  the change of the shear
and bulk viscosities with thermal parameters obtained in this paper may depend on specific model dynamics.
Nevertheless, one  expects that some of our  results,  e.g. that on the influence of the order of the chiral phase transition on $\eta$ and $\zeta$, could be relevant for qualitative  understanding of transport  properties in the QCD medium.

The paper is organized as follows:  In Section \ref{sec:njl} we introduce  the NJL model and its thermodynamics. In Section  \ref{sec:coeff} we describe the methods used to calculate the shear and bulk viscosity.  We also quantify in this section the properties of the shear and bulk viscosity near the phase transition and along  the phase boundary. Finally, in Section \ref{sec:sum} we summarize  our results.

\section{The two-flavor NJL model}
\label{sec:njl}

In order to quantify the transport properties of fluid composed of constituent quarks  near the chiral
phase transition we use the Nambu--Jona-Lasinio (NJL) model~\cite{nambu}. This model exhibits basic
properties expected in QCD due to the chiral symmetry restoration. 
We start with the following Lagrangian for two quark flavors
and  three colors~\cite{review}
\begin{eqnarray}
{\mathcal L}
&=& \bar{\psi}( i\Slash{\partial} -m )\psi
{}+ \bar{\psi}\mu\gamma_0\psi
\nonumber\\
&&
{}+ G_S \left[ \left( \bar{\psi}\psi \right)^2
{}+ \left( \bar{\psi}i\vec{\tau}\gamma_5\psi \right)^2
\right]\,,
\end{eqnarray}
where $m = \mbox{diag}(m_u, m_d)$  and  $\mu = \mbox{diag}(\mu_u, \mu_d)$ are   the current quark
 masses and the  quark chemical potentials respectively, whereas $\vec{\tau}$ are Pauli matrices.
The coupling of  four-fermion interactions $G_S \Lambda^2 = 2.44$ and the  three momentum cut-off
$\Lambda=587.9$ MeV are chosen so as
 to reproduce  the vacuum pion decay constant and chiral condensate  with $m_u=m_d=5.6$ MeV.

In the mean field approximation   and for the isospin symmetric system the thermodynamics of the NJL
 model is described by the
potential:
\begin{eqnarray}
&&
\Omega (T,\mu;M)/V = \frac{(M-m)^2}{4G_S}
{}- 12 \int\frac{d^3p}{(2\pi)^3}
\left[E(\vec{p}\,)
\right.
\nonumber\\
&&
\left.
{}- T\ln ( 1-f(\vec{p},T,\mu))
{}- T\ln (1-\bar{f}(\vec{p},T,\mu) \right]\,,
\label{omega}
\end{eqnarray}
with  $M = m- 2G_S\langle \bar{\psi}\psi \rangle$ being a dynamical quark mass, $E(\vec{p}\,) =
\sqrt{\vec{p}^{\,2} + M^2}$  the quasi-particle energy and $f,\bar{f} = \Bigl( 1 + \exp\bigl[
(E(\vec{p}\,) \mp \mu)/T \bigr] \Bigr)^{-1}$ are the particle/antiparticle distribution functions.
The dynamical quark mass $M$ in Eq.~(\ref{omega}) is obtained self-consistently from the
stationarity condition ${\partial\Omega}/{\partial M} = 0$ that leads to:
\begin{equation}
M = m + 12 G_S \int\frac{d^3 p}{(2\pi)^3}
\frac{M}{E} \left[ 1 - f - \bar{f} \right]\,.
\label{gapeq}
\end{equation}

\begin{figure}
\begin{center}
\includegraphics[width=8cm]{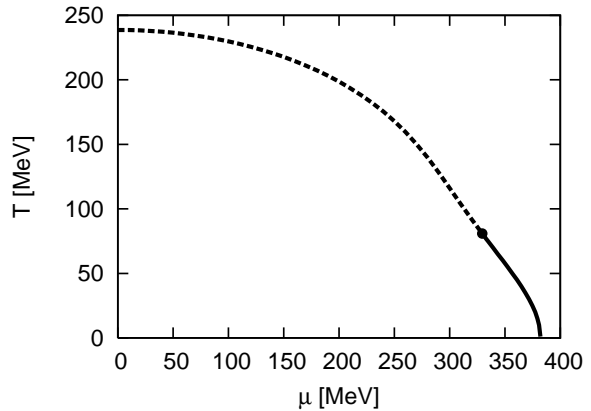}
\caption{ The phase diagram of the NJL model.
The full and dashed lines show the first order 
and cross over transition respectively. The dot indicates the position of the CEP
located at $(T,\mu)=(80.9,329)$ MeV.
}
\label{phase}
\end{center}
\end{figure}

The thermodynamic  potential of the  NJL model exhibits  a typical QCD--like phase diagram shown in
Fig.~\ref{phase}. The cross-over transition is identified from the peak position of the chiral
susceptibility
\begin{equation}
\chi_{\rm ch} = - \frac{\partial\langle\bar{\psi}\psi\rangle}
{\partial m}\,.
\end{equation}
The CEP is obtained from  the condition of the zero-curvature of an effective
potential or equivalently from  the divergence of the quark-number susceptibility.

Our main  goal is to  quantify the transport properties of a medium composed of  dynamical quarks with
mass $M(T,\mu)$  near the chiral phase transition and along the boundary line shown in the Fig.
(\ref{phase}). In this context we  explore
 the shear $\eta$ and bulk $\zeta$ viscosities.

 In the thermodynamic system $\eta$ and $\zeta$ are sensitive to thermal and dynamical properties
of a medium. To explicitly calculate transport coefficients we  use the approach based on the
relativistic kinetic theory.

\section{Transport coefficients}
\label{sec:coeff} 

In the relativistic kinetic theory the transport parameters, 
the shear   and the bulk viscosities,  are defined as coefficients 
of the space--space component of  the energy-momentum tensor 
away from equilibrium.

For the fluid composed of quasi-particles with dynamical mass $M(T,\mu)$ and under the relaxation
time approximation, the shear and bulk viscosities  are obtained as follows~\cite{sr}
\begin{widetext}
\begin{eqnarray}\label{eq_sr}
\eta
&=& \frac{4}{5T}\int\frac{d^3p}{(2\pi)^3}
\frac{\vec{p}^4}{E^2}\left[ \tau f(1-f)
{}+ \bar{\tau}\bar{f}(1-\bar{f})
\right]\,,
\label{shear}
\\
\zeta
&=& -\frac{4}{T}\int\frac{d^3p}{(2\pi)^3}
\frac{M^2}{E}
\left[
\left( \tau f(1-f) + \bar{\tau}\bar{f}(1-\bar{f})
\right)
\left( \frac{\vec{p}^2}{3E}
{}- \left(\frac{\partial P}{\partial \epsilon} \right)_n
\left( E - T\frac{\partial E}{\partial T}
{}- \mu \frac{\partial E}{\partial\mu} \right)
{}+ \left(\frac{\partial P}{\partial n}\right)_\epsilon
\frac{\partial E}{\partial\mu}
\right)
\right.
\nonumber\\
&&
\left.
{}- \left( \tau f(1-f) - \bar{\tau}\bar{f}(1-\bar{f})
\right) \left(\frac{\partial P}{\partial n}\right)_\epsilon
\right]\,.
\label{bulk}
\end{eqnarray}
\end{widetext}
The derivatives of pressure $P$ with respect to the net quark number density $n$ or energy density
$\epsilon$ in Eqs. (\ref{shear}) and  (\ref{bulk}) can be expressed in terms of  susceptibilities
$\chi_{xy}= \partial^2P/\partial x \partial y$ as
\begin{eqnarray}
\left( \frac{\partial P}{\partial \epsilon}\right)_n
&=& \frac{s\chi_{\mu\mu} - n \chi_{\mu T}}
{C_V \chi_{\mu\mu}}\,,
\nonumber\\
\left( \frac{\partial P}{\partial n}\right)_\epsilon
&=&  \frac{nT\chi_{TT} + (n\mu - sT)\chi_{\mu T}
{}- s\mu \chi_{\mu\mu}}
{C_V \chi_{\mu\mu}}\,,
\end{eqnarray}
with $C_V$ being the specific heat calculated  at a constant volume.  The $C_V$ is also expressed
through susceptibilities
\begin{equation}
C_V = T \left(\frac{\partial s}{\partial T} \right)_V
= T \left[ \chi_{TT} - \frac{\chi_{\mu
T}^2}{\chi_{\mu\mu}} \right]\,.
\end{equation}

To quantify the change of the above transport coefficients in quasi--particle  medium one still
needs to specify  the relaxation time $\tau$. The $\tau$ is calculated from  the thermal averaged
cross-section describing the total elastic scattering of medium constituents. In the NJL model the
medium is composed from constituent quarks with two different flavors. Thus,  in the dilute gas
approximation the relaxation time $\tau$  is obtained from  \cite{reif,DG}
\begin{eqnarray}
\tau^{-1}
&=&
6 \left[ n_q \left( \bar{\sigma}_{uu \to uu}
{}+ \bar{\sigma}_{ud \to ud} \right)
\right.
\nonumber\\
&&
\left.
{}+ n_{\bar{q}} \left( \bar{\sigma}_{u\bar{u} \to u\bar{u}}
{}+ \bar{\sigma}_{u\bar{u} \to d\bar{d}}
{}+ \bar{\sigma}_{u\bar{d} \to u\bar{d}}\right)
\right]\,,
\end{eqnarray}
where $n_i$ are the quark/antiquark densities
\begin{eqnarray}
n_q
&=&
\int \frac{d^3 p}{(2\pi)^3} f(p;T,\mu)\,,
\nonumber\\
n_{\bar{q}}
&=&
\int \frac{d^3 p}{(2\pi)^3} \bar{f}(p;T,\mu)\,.
\end{eqnarray}

\begin{figure}
\begin{center}
\includegraphics[width=8cm]{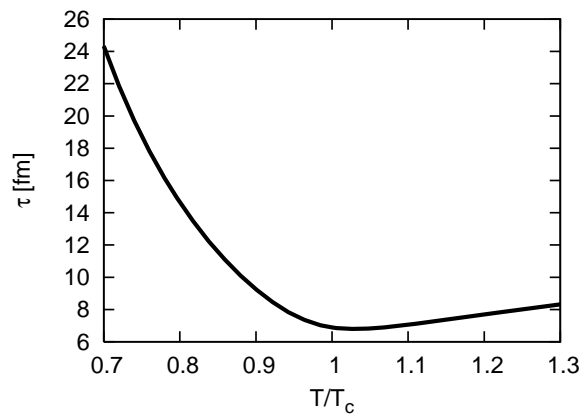}
\caption{The collision time in the NJL model at $\mu=0$.
}
\label{taumu0}
\end{center}
\end{figure}

The in--medium cross section  for quark-antiquark, quark-quark and antiquark-antiquark scattering
processes were studied in detail in Ref. \cite{eta:njl} including    $1/N_c$ next to leading order
corrections. For $u\bar{u} \to u\bar{u}$ scattering   the differential cross section is obtained
from the T matrix as
\begin{eqnarray}\label{dif}
&&
\frac{d\sigma_{u\bar{u} \to u\bar{u}}}{dt}(s,t;T,\mu)
\nonumber\\
&&
= \frac{1}{16\pi s (s - 4M^2)}
\sum_{c,s}\left| T_{u\bar{u} \to u\bar{u}} \right|^2
(s,t;T,\mu)\,,
\end{eqnarray}
where the average over the initial color and spin as well as  the summation over final states were
taken. We consider  the rest frame of colliding quarks so that the differential cross section
(\ref{dif}) is expressed as the function of the Mandelstam variables $s$ and $t$. The integrated cross
section is calculated from
\begin{eqnarray}
&&
\sigma_{u\bar{u} \to u\bar{u}}(s;T,\mu)
= \int dt \frac{d\sigma_{u\bar{u} \to u\bar{u}}}{dt}
\frac{-4t (s + t - 4M^2)}{(s - 4M^2)^2}
\nonumber\\
&&
\times
\left( 1 - f(\sqrt{s}/2 - \mu) \right)
\left( 1 - \bar{f}(\sqrt{s}/2 + \mu) \right)\,.
\end{eqnarray}
In the above equation  we have taken into account  that the transport process  is dominated by  the
large angle scattering ~\cite{reif}. The terms in the  brackets correspond to the Pauli blocking
factors which account for a possible occupation of particles in the final state. We also
introduce    the energy-averaged cross section
\begin{equation}
\bar{\sigma}_{u\bar{u} \to u\bar{u}}(T,\mu)
= \int ds \sigma_{u\bar{u} \to u\bar{u}}(s;T,\mu)P(s;T,\mu)\,,
\end{equation}
with  $P$ being  the probability of yielding  $q\bar{q}$ pair with the energy $s$
\begin{eqnarray}
&&
P(s;T,\mu) = C \sqrt{s (s - 4M^2)}
\nonumber\\
&&
\quad
\times
f(\sqrt{s}/2 - \mu)\,
\bar{f}(\sqrt{s}/2 + \mu)\,v_{\rm rel}(s)\,,
\end{eqnarray}
where  the relative velocity between two particles in the initial state $v_{\rm rel}$ is given by
\begin{equation}
v_{\rm rel}(s) = \sqrt{\frac{s - 4M^2}{s}}\,.
\end{equation}
The normalization constant $C$ is fixed from
\begin{equation}
\int ds P(s;T,\mu) = 1\,.
\end{equation}
The differential cross sections for different   quark flavors are obtained through the  isospin
symmetry, charge conjugation and crossing symmetry. The total cross section after the energy
averaging reads
\begin{equation}\label{total}
\bar{\sigma}_q = \bar{\sigma}_{u\bar{u}\to u\bar{u}}
{}+ \bar{\sigma}_{u\bar{u}\to d\bar{d}}
{}+ \bar{\sigma}_{u\bar{d}\to u\bar{d}}
{}+ \bar{\sigma}_{uu \to uu}
{}+ \bar{\sigma}_{ud \to ud}\,.
\end{equation}
The dominant contribution to the total   cross section (\ref{total}) comes from the s-channel due
to the propagation of $\pi$ and $\sigma$ modes.
 The mass of $\sigma$ drops toward the critical temperature $T_c$ and due to chiral symmetry
restoration  is degenerated with the pion mass at $T=T_c$. On the other hand, for $T>T_c$, both   $\pi$
and $\sigma$ masses are increasing with temperature. Consequently, the cross section
$\sigma(s=m_\sigma^2;T,\mu)$ shows  singularity
at $T_c$~\cite{eta:njl}~\footnote{%
 More precisely, the singularity appears at the pionic Mott temperature at which the pion mass
 becomes equal to the mass of its constituents, $m_\pi = 2M$.  In the chiral limit the Mott
 temperature coincides with the chiral phase transition temperature.
}. However, after the integration over energy $s$  this singularity is mostly  washed out in the
total cross section $\bar{\sigma}_q$. As remnants of the above singularity a broad bump in  the
$\bar{\sigma}_q$  is seen arround  $T_c$~\cite{eta:njl}. Thus, the total cross section is
non-singular even at the CEP.

\begin{figure*}
\begin{center}
\includegraphics[width=8.7cm]{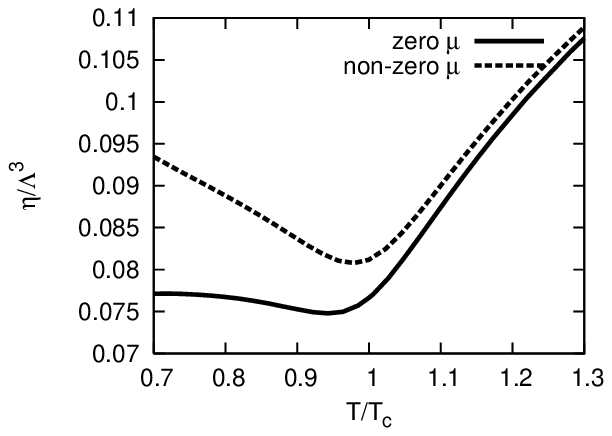}
\includegraphics[width=8.7cm]{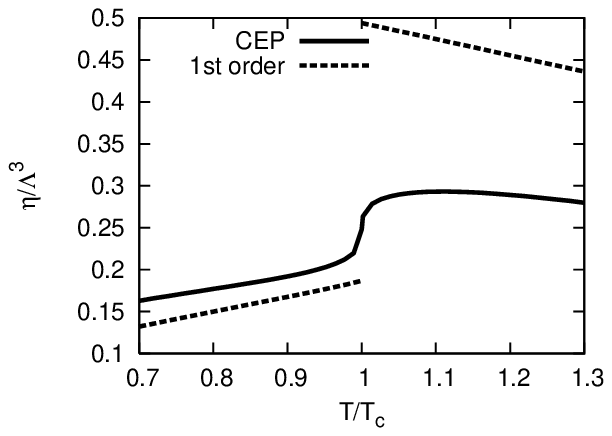}
\caption{The shear viscosity normalized by the 3-momentum cutoff around cross over (left--hand figure) and
around $\mu_{\rm CEP}$ and first order transition (right--hand figure). } \label{etamu}
\end{center}
\end{figure*}

\begin{figure*}
\begin{center}
\includegraphics[width=8.7cm]{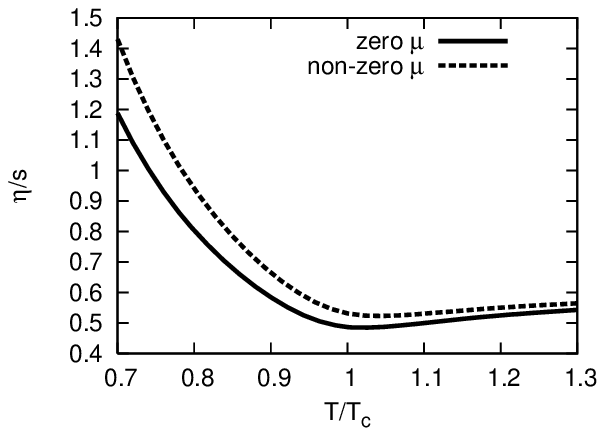}
\includegraphics[width=8.7cm]{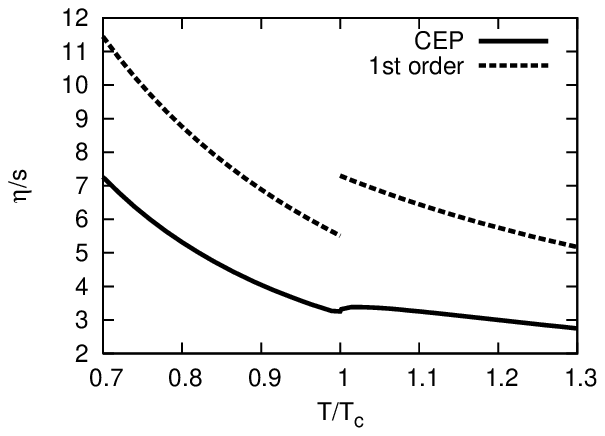}
\caption{The ratio of shear viscosity to entropy density around cross over (left--hand figure) and around
$\mu_{\rm CEP}$ and first order transition (right--hand figure). } \label{esmu}
\end{center}
\end{figure*}

Fig.~\ref{taumu0} shows the collision time calculated from Eq. (\ref{eq_sr}) at finite  $T$ and at  $\mu
= 0$. There is a clear decrease of $\tau$ when approaching the critical region from the chirally broken
phase. One should realize that
 in spite of the phase transition  the collision time is quite smooth around  $T_c$ and
experiences only a shallow minimum at the phase transition.
 This behavior is due to the total cross section which shows a similar broad
structure around $T_c$. The $\tau$ is large at low and small at high temperatures since the quark
density is increasing
function of temperature ~\footnote{%
 The behavior of $\tau$ at $T \ll T_c$  seen in Fig. \ref{taumu0} is specific  for  the NJL model
  where in the low-temperature phase the  medium is  composed only of  heavy constituent   quarks.
}.

At the finite $\mu$ the total cross--section is larger than that at $\mu = 0$  and the   quark density
increases while the anti-quark density decreases. Consequently, the collision time for  quark
$\tau$ and for  anti-quark $\bar\tau$ behaves as
\begin{equation}
\tau(\mu) > \tau(\mu=0)\,,
\quad
\bar{\tau}(\mu) < \bar{\tau}(\mu=0)\,.
\end{equation}

With the above results for the collision time applied in Eq. (\ref{eq_sr}) we can calculate  the
transport coefficients in the quasi-quark medium and study their sensitivity to the chiral     phase
transition.


\subsection{Bulk and shear viscosities near phase transition}
\label{ssec:etamu}

The temperature and density dependence of the shear viscosity obtained in the NJL model under
relaxation time approximation is shown in Fig.~\ref{etamu}. For low density the shear viscosity
shows a  shallow minimum near the pseudo-critical temperature. This behavior is   similar as that
observed in  the collision time. The constituent quark mass decreases with temperature and the
quark density is  enhanced in the chirally symmetric phase which   results in the rise of $\eta$ at
high temperature.  Quantitatively the difference in $\eta$ at $\mu=0$ and $\mu \neq 0$ comes mainly
from the density effect. Consequently, at the CEP and near  the  first-order phase transition, this
effect  should be even more significant. Fig.~\ref{etamu}--(right) shows the change of shear
viscosity when crossing the CEP and the boundary of the first order phase transition. As expected,
the $\eta$ shows discontinuity at  the first order transition. At the CEP this discontinuity
disappears, however there is still an  abrupt change of $\eta$. One observes that the
shear viscosity is decreasing toward the CEP and first order phase transition from the side of the symmetric phase.

When discussing the transport properties of  relativistic fluid a natural normalization of shear
viscosity is the entropy density $s$. In this way the $\eta/ s$ is dimensionless quantity that
characterizes dissipation of energy in a medium.
Fig.~\ref{esmu} shows the temperature dependence of the ratio $\eta/s$ at  vanishing  and at finite
chemical potential. The entropy density rapidly increases with temperature, thus  the $\eta/s$
is reduced in the high-temperature phase. A shallow minimum structure around $T_c$ for  small
$\mu$ is not preserved  with increasing $\mu$. From Fig.~\ref{esmu}--(right) one sees that $\eta/s$
is insensitive to the existence of the CEP. At higher $\mu$ the ratio shows a jump associated with
the first order phase transition.


\begin{figure*}
\begin{center}
\includegraphics[width=8.7cm]{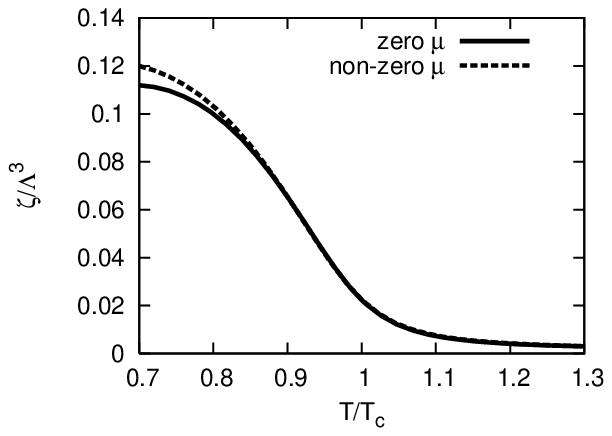}
\includegraphics[width=8.7cm]{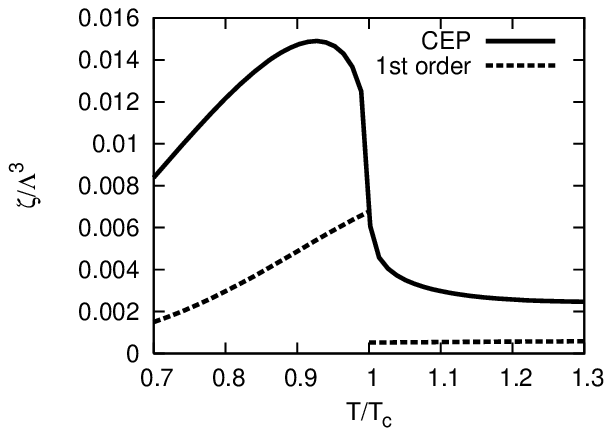}
\caption{The bulk viscosity normalized by the 3-momentum cutoff around the cross over (left--hand figure) and around
$\mu_{\rm CEP}$ and first order transition (right--hand figure). } \label{zetamu}
\end{center}
\end{figure*}

\begin{figure*}
\begin{center}
\includegraphics[width=8.7cm]{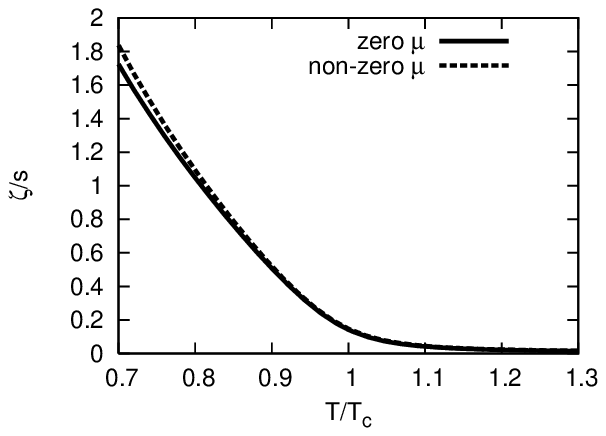}
\includegraphics[width=8.7cm]{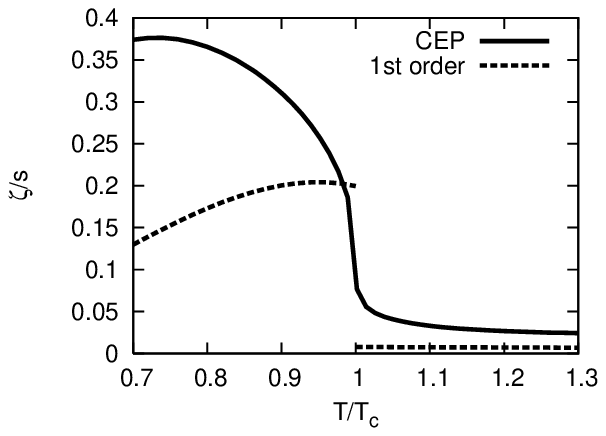}
\caption{The ratio of bulk viscosity to entropy density around the cross over (left--hand figure) and around
$\mu_{\rm CEP}$ and first order transition (right--hand figure). } \label{zsmu}
\end{center}
\end{figure*}

From the general analytic expression for the shear  and the bulk  
viscosities  (see Eqs. (\ref{bulk}) and (\ref{shear})) one  expects  that  $\zeta$ should be
 more sensitive to the phase transition than  $\eta$ due to explicit contribution of derivative terms involving a dynamical quark mass.  Fig.~\ref{zetamu} shows the temperature dependence of $\zeta$
 for different $\mu$. In the regime of cross--over transition the density dependence of $\zeta$ is
 much weaker  than that seen in the shear viscosity. This is indeed due  to $\partial M/\partial T$ and $\partial M/\partial\mu$
 terms that contribute to the bulk  and are absent  in the shear  viscosity. The temperature
 dependence of $\zeta$ at finite  $\mu$ is 
similar to that at $\mu = 0$ as seen  in Fig.~\ref{zetamu}-(left). With the increasing chemical
potential, the quark density effects start to be visible. The growth of $\zeta$ toward $T_c$ in
Fig.~\ref{zetamu}-(right) is due to an  increase of quark number density. At the CEP the bulk
viscosity is finite in spite of the fact that all susceptibilities contributing to $\zeta$ are
diverging  \cite{sr}. Thus, at the CEP there is a precise cancelation of singularities from different terms in Eq.
(\ref{bulk}). Nevertheless, one observes a rapid downward change approaching the CEP from
the low--temperature side. This reflects the change of the dynamical quark mass with  temperature.

The ratio   $\zeta/s$, of the bulk viscosity to the entropy density, keeps basically a  structure of
$\zeta/\Lambda^3$. The suppressed entropy density due to the  heavy constituent quark mass in
low-temperature phase results in  growth of the ratio shown in Fig.~\ref{zsmu}. The rate of change
in  $\zeta/s$ with  temperature at CEP is higher than that of $\eta/s$. However, the  absolute
value of $\zeta/s$  is smaller than $\eta/s$   at CEP. This result is valid under relaxation
time approximation where $\tau$ is always finite. However, when going beyond this approximation and
including dynamical scaling  \cite{onuki}, the bulk viscosity should be  singular at the
CEP.

\begin{figure*}
\begin{center}
\includegraphics[width=8.7cm]{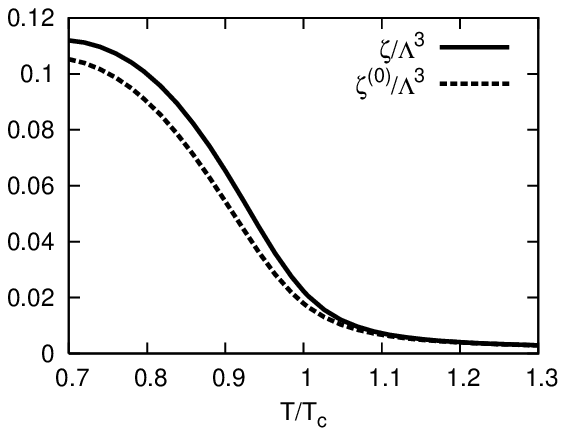}
\includegraphics[width=8.7cm]{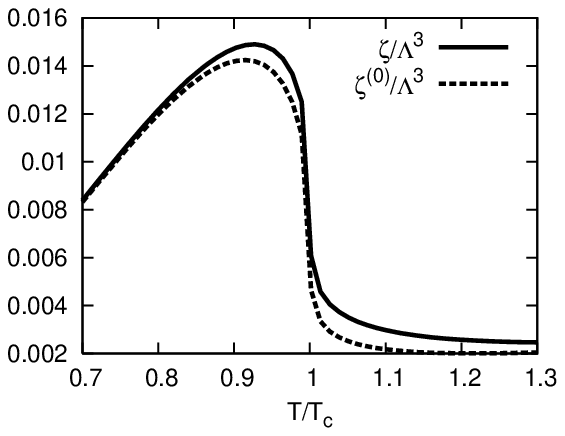}
\caption{The bulk viscosity normalized by the 3-momentum cutoff at $\mu=0$ (left--hand figure) and at $\mu_{\rm
CEP}$ (right--hand figure). The solid line is obtained from Eq.~(\ref{bulk}) whereas  the dashed line  from the same equation after  eliminating the terms proportional  to mass derivatives.  } \label{noder}
\end{center}
\end{figure*}

In the derivation of the  bulk viscosity  we have explicitly included in Eq. (\ref{bulk}) the modification of
particle dispersion relations due to a dynamically generated  quasi-particle mass  which is $T$ and
$\mu$ dependent. In Fig.~\ref{noder} we quantify the influence of  derivatives  of a dynamical mass
on the bulk viscosity. The full--line in this figure was obtained  from  Eq.~(\ref{bulk}) whereas the
dashed--line   was calculated by eliminating
 the $\partial M/\partial T$ and $\partial M/\partial\mu$ terms  from Eq.~(\ref{bulk}). The behavior of $\zeta$ is largely governed by the
dynamical quark mass $M(T,\mu)$ multiplied in front of (\ref{bulk}). Unlike  the expectation that
the bulk viscosity should be  sensitive to the phase transition, there is no much difference in the
above two cases. The  enhancement of $\partial M/\partial T$ around $T_c$  is weakened by the pre-factor $M^2$ in Eq.~(\ref{bulk}).  In addition, the
contributions of different susceptibilities in Eq.~(\ref{bulk}) which are diverging at the CEP
are canceled out, leading to a finite bulk viscosity at the critical end point ~\cite{sr}.

\begin{figure*}
\begin{center}
\includegraphics[width=8.7cm]{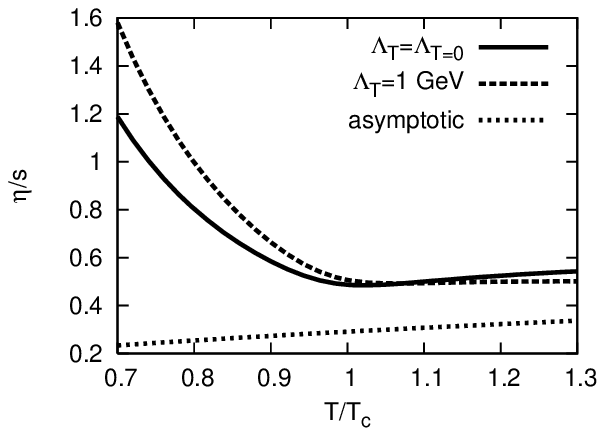}
\includegraphics[width=8.7cm]{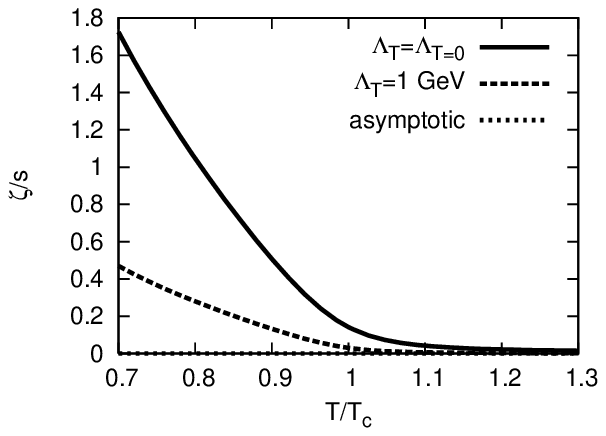}
\caption{The ratio of $\eta/s$ (left--hand figure) and $\zeta/s$ (right--hand figure) at $\mu=0$. The solid curves are
obtained  with  the same 3-momentum cutoff $\Lambda_T$ in thermal part as in the vacuum, i.e.,
$\Lambda_T = 587.9$ MeV. The dashed lines are obtained for $\Lambda_T = 1$ GeV in themal part while
in vacuum part $\Lambda = 587.9$ MeV is used. The dotted curves indicate the asymptotic values for
massless quarks in high temperature limit. } \label{escut}
\end{center}
\end{figure*}

The NJL model, being  non-renormalizable, requires the ultraviolet momentum cut-off $\Lambda$. Consequently,
all thermodynamic quantities are affected by finite $\Lambda$. This is particularly the case in the
parameter range   where the physics is sensitive to  a large particle momenta. Fig.~\ref{escut} shows
sensitivity of  $\eta/s$ and $\zeta/s$ to the momentum cut-off $\Lambda_T$ implemented in the
thermal part of thermodynamic potential. We compare the results obtained with
$\Lambda_T=\Lambda_{T=0}=587.9$ MeV  and with $\Lambda_T$=1 GeV. In  high temperature phase we also
show in Fig.~\ref{escut}
 the resulting    viscosities obtained from the perturbative QCD 
(pQCD)~\cite{HK}~\footnote{For the recent results of viscosity parameters obtained in the high temperature QCD see Ref. \cite{arnold}.}
\begin{eqnarray}
&&
\eta^{\rm (asymptotic)}
= \frac{4\pi^2}{675}\frac{3.86}{\alpha_s^2\ln(1/\alpha_s)}
T^3\,,
\nonumber\\
&&
\zeta^{\rm (asymptotic)} = 0\,,
\end{eqnarray}
with  the two-loop running coupling for $N_f=2$
\begin{equation}
\alpha_s = 2\pi \left[ \frac{29}{3}\ln(T/\lambda)
{}+ \frac{115}{29}\ln(2\ln(T/\lambda)) \right]^{-1}\,,
\end{equation}
and with $\lambda = 30$ MeV.

\begin{figure*}
\begin{center}
\includegraphics[width=8.7cm]{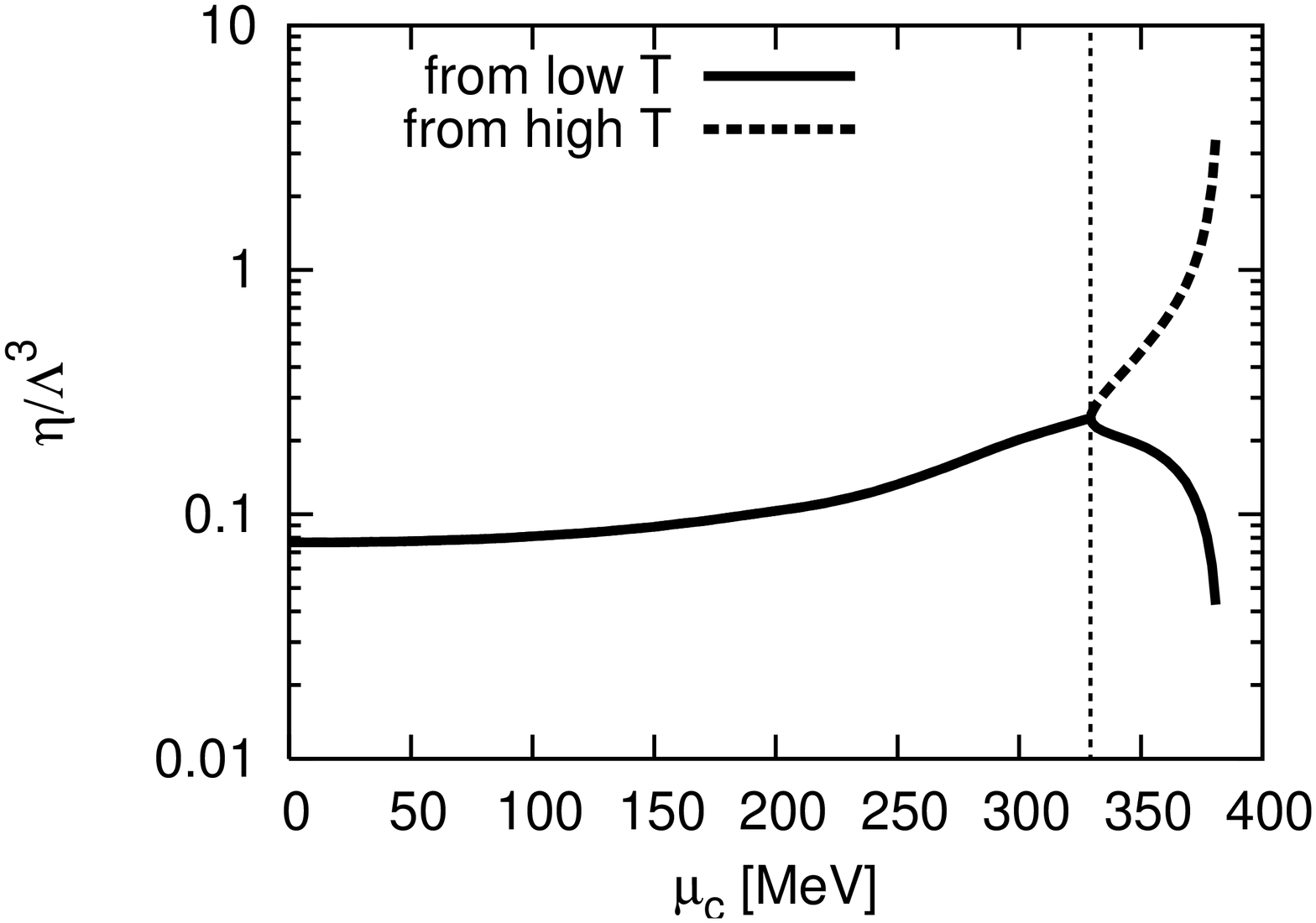}
\includegraphics[width=8.7cm]{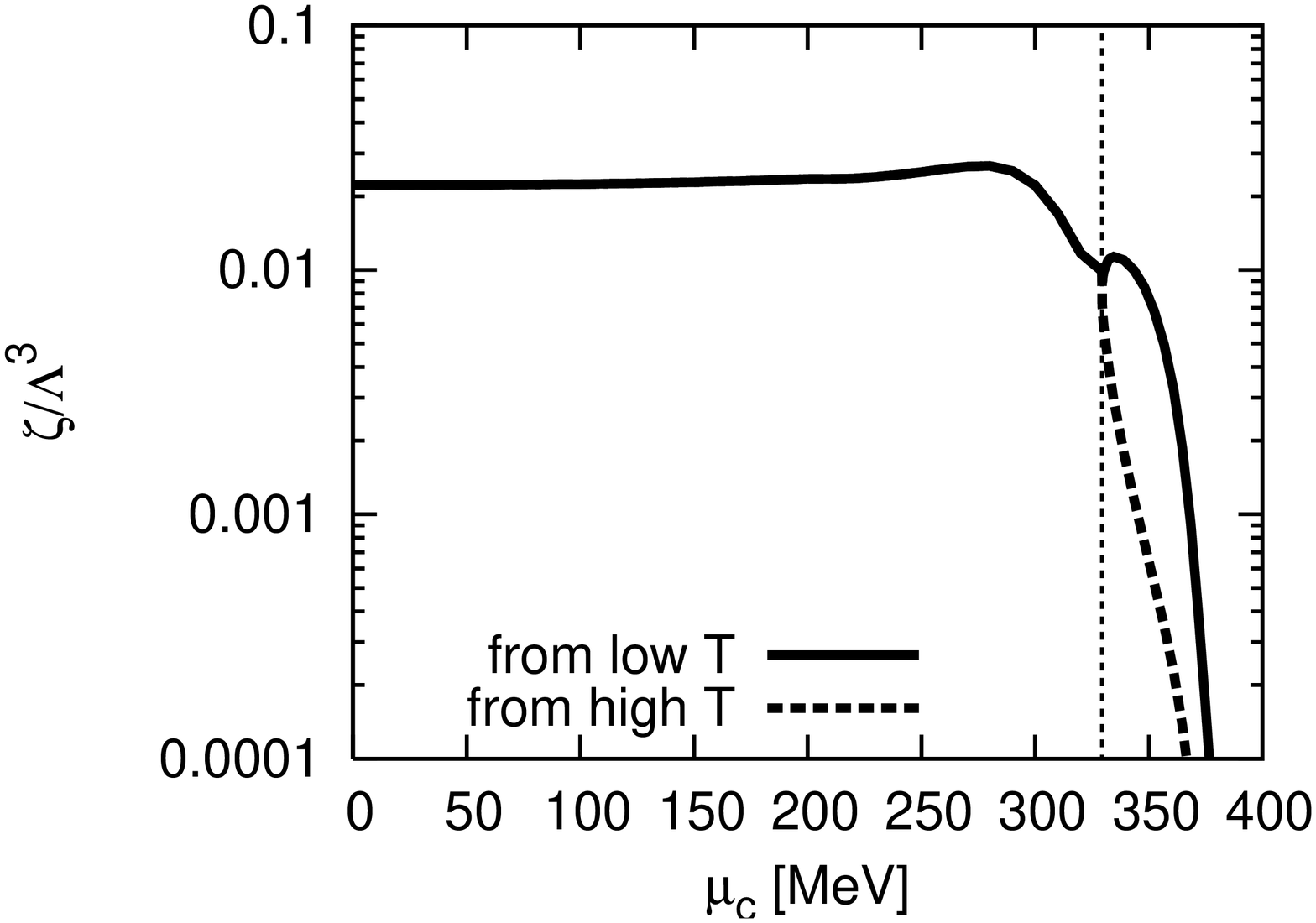}
\caption{The shear (left--hand figure) and bulk (right--hand figure) viscosities normalized by the 3-momentum cutoff
calculated along the phase boundary. The vertical dotted-line indicates the position of the CEP. } \label{etacl}
\end{center}
\end{figure*}
\begin{figure*}
\begin{center}
\includegraphics[width=8.7cm]{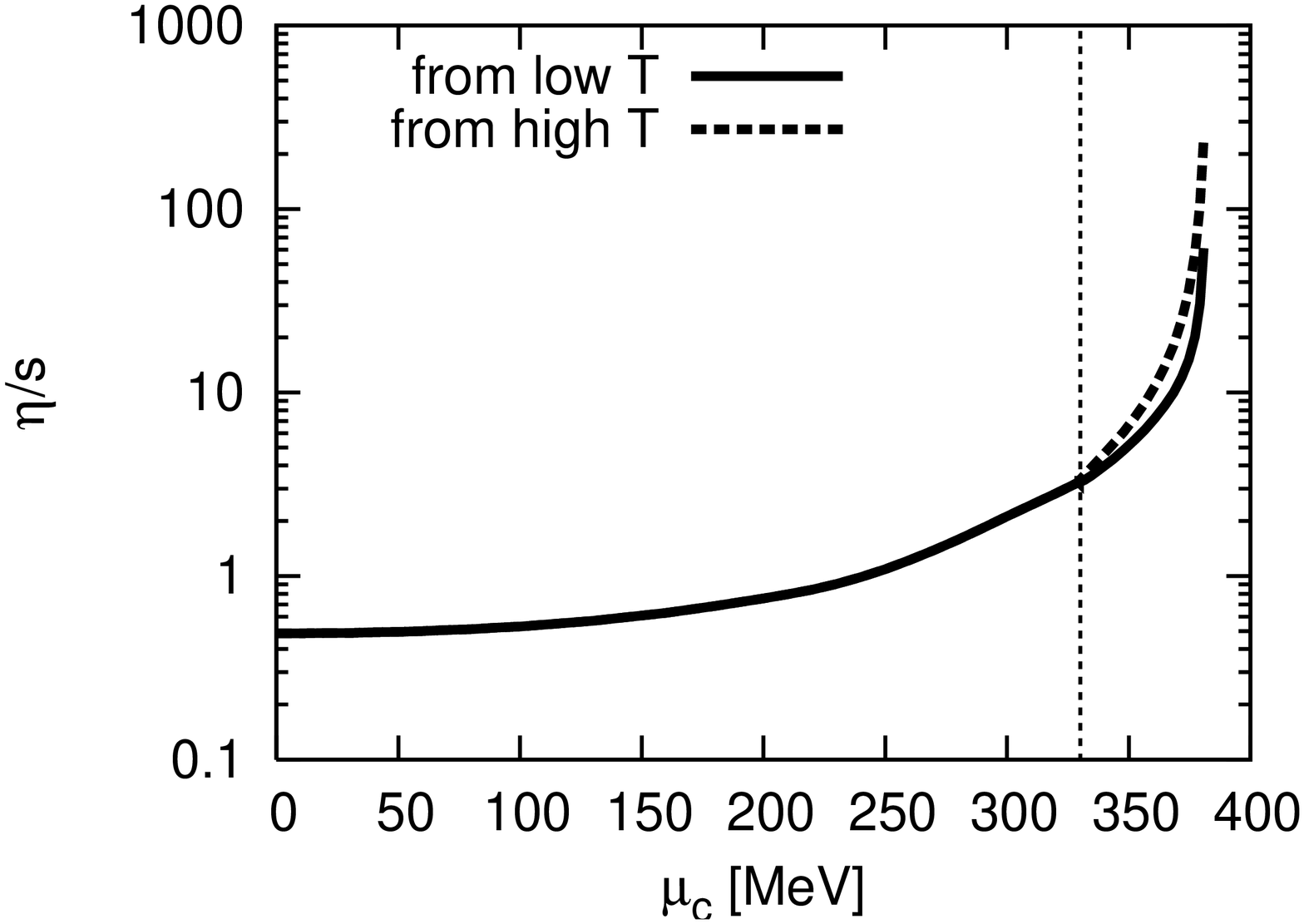}
\includegraphics[width=8.7cm]{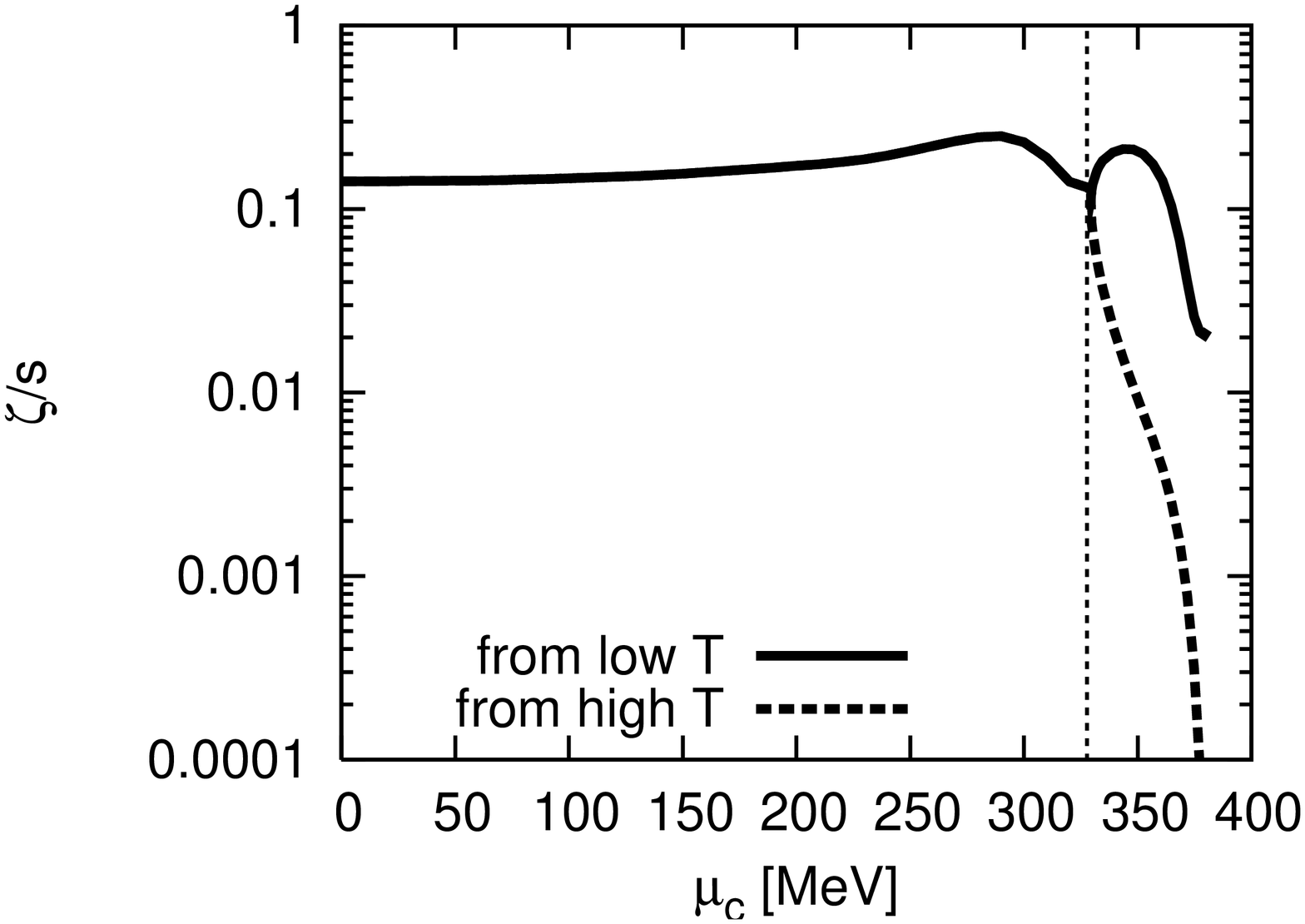}
\caption{The ratio of shear (left--hand figure) and bulk (right--hand figure) viscosities to entropy density calculated along
the phase boundary. The vertical dotted-line indicates the position of the CEP. } \label{escl}
\end{center}
\end{figure*}

As seen in Fig.~\ref{escut}, both $\eta/s$ and $\zeta/s$ are very sensitive to the momentum cut-off.
However, at low $T$ this dependence is opposite in shear and bulk viscosity. There is an increase
of $\eta/s$ and decrease of
 $\zeta/s$ with increasing $\Lambda$. This is due to derivative terms in the bulk viscosity which
   show  steeper change with $\Lambda$. At high temperature     the NJL model  results  tend to
   converge to that expected in pQCD. The bulk viscosity
 vanishes at high temperature, as expected in the conformal theory. However, due to the absence of gluons and the presence of the finite cut-off  the
 shear viscosity deviates from their  asymptotic  value even at $T \gg T_c$.

\subsection{Viscosities along the   phase boundary}
\label{ssec:etacl}

Fig.~\ref{etacl} shows the shear and bulk viscosities calculated along the  chiral phase boundary  obtained in the NJL model and shown in Fig. \ref{phase}.
Two lines at high $\mu_c$ indicate a gap associated with the first order transition. The shear
viscosity increases along the cross--over line towards  the CEP where it reaches its maximal value.
The bulk viscosity is rather weekly changing with $T$ in the regime of cross--over and exhibits a local minimum at the CEP.  
A little bump on the cross-over line before reaching the CEP
is an artifact of the present NJL model calculation. In a more
realistic calculation, this may disappear and the bulk viscosity
smoothly goes down to the CEP.
The $\eta$ shows  an upward jump approaching the first order transition
from the low-temperature phase. This is an opposite behavior to $\zeta$ which  shows a downward jump.

Fig.~\ref{escl} shows the ratios, $\eta/s$ and $\zeta/s$, calculated along the critical line. 
The $\eta/s$ is monotonic along the critical line without showing any sensitivity to the  CEP. The
 $\zeta/s$  preserves the local minimum at the CEP and shows a visible gap at the first order
transition.

The properties  of the transport coefficients  calculated within  NJL model are not  universal in
spite of the fact that this model belongs to the  QCD universality class with respect  to chiral
symmetry. In  addition the shear and bulk viscosities are sensitive to particle composition of the
medium and  to specific model dynamics that enters to various collision processes.   The shear and
bulk viscosities   do not show   critical behavior with static critical exponents. Nevertheless,
one expects that some generic futures of transport coefficients along the phase boundary calculated
within  the above model could be of interest for understanding transport properties of  the  QCD
matter.
From this perspective, a trace of in-medium change of the order 
parameter $M(T,\mu)$ remains more distinctly in the bulk
than in the shear viscosity.

\section{Summary and conclusions}
\label{sec:sum}

Based on the NJL model  and kinetic theory we have shown the transport properties of thermal
 medium composed of dynamical quarks. We have included in our calculations in-medium modification
 of constituent quark dispersion relations.
We have analyzed  behaviors of shear $\eta$  and bulk $\zeta$ viscosities near the chiral phase transition
within the mean field approximation. The collision time required to quantify  viscosity parameters
was calculated from the elastic scattering cross-section of constituent quarks
including next to leading $1/N_c$ corrections.

We have shown that around the phase transition the shear viscosity exhibits only a shallow minimum
for small chemical potentials. For higher $\mu$ corresponding to the  first order transition and near the
critical end point the shear viscosity is  discontinuous or exhibits
 an  abrupt change. However, when normalizing $\eta$ by  the entropy density $s$ this abrupt
change at  CEP disappears. The bulk viscosity is more sensitive to the change of the dynamical
quark mass around the phase transition, thus shows steeper behavior near $T_c$, however stays  non-singular at CEP.

Along the chiral phase boundary  the $\eta/s$ is changing monotonically with critical parameters
whereas the $\zeta/s$ has a local minimum at the CEP. Nevertheless,
 within the considered  model  and under relaxation time approximations none  of viscosities could
be considered as a useful  probe of the critical end point in the phase diagram. We have to stress
that due to the  non--universal behavior of bulk and shear
 viscosity within  these calculations  our quantitative results  are  model dependent.
 However, we expect that qualitative behavior of $\eta/s$ and $\zeta/s$ near  the chiral phase
 boundary derived  here could be less sensitive to  the specific model dynamics since important change of the order paratmer is certainly incorporated in our formula.

\section*{ Acknowledgments}
We acknowledge interesting discussions with B. Friman and J. Wambach.  C.S. acknowledges partial support by the DFG
cluster of excellence ``Origin and Structure of the Universe''. K.R. acknowledges partial support
of the Polish Ministry of Science and Higher  Education (MENiSW) and  the Deutsche Forschungsgemeinschaft  (DFG) under the "Mercator program".



\begin{thebibliography}{99}

\bibitem{hydro} J.~Kapusta,
  Phys.\ Rev.\  C {\bf 24}, 2545 (1981). M.~Prakash, M.~Prakash, R.~Venugopalan and G.~Welke,
  Phys.\ Rept.\  {\bf 227}, 321 (1993). P.~Kovtun, D.~T.~Son and A.~O.~Starinets,
  Phys.\ Rev.\ Lett.\  {\bf 94}, 111601 (2005). T.~Hirano and M.~Gyulassy,
  Nucl.\ Phys.\  A {\bf 769}, 71 (2006). P.~Romatschke and U.~Romatschke,
  Phys.\ Rev.\ Lett.\  {\bf 99}, 172301 (2007). A.~Muronga,
  Phys.\ Rev.\  C {\bf 76}, 014909 (2007); Phys.\ Rev.\  C {\bf 76}, 014910 (2007).
K.~Dusling and D.~Teaney, Phys.\ Rev.\  C {\bf 77}, 034905 (2008).
H.~Song and U.~W.~Heinz, Phys.\ Lett.\  B {\bf 658}, 279 (2008).
G.~Torrieri, B.~Tomasik and I.~Mishustin, Phys.\ Rev.\  C {\bf 77}, 034903 (2008).

\bibitem{kapusta}
  J.~I.~Kapusta,
  arXiv:0809.3746 [nucl-th].


\bibitem{eta:njl} P.~Zhuang, J.~Hufner, S.~P.~Klevansky and L.~Neise,
  Phys.\ Rev.\  D {\bf 51}, 3728 (1995). P.~Rehberg, S.~P.~Klevansky and J.~Hufner,
  Nucl.\ Phys.\  A {\bf 608}, 356 (1996).

\bibitem{hadronic} A.~Dobado and F.~J.~Llanes-Estrada,
  Phys.\ Rev.\  D {\bf 69}, 116004 (2004). J.~W.~Chen, Y.~H.~Li, Y.~F.~Liu and E.~Nakano,
  Phys.\ Rev.\  D {\bf 76}, 114011 (2007). P.~Czerski, W.~M.~Alberico, S.~Chiacchiera, A.~De Pace,
H.~Hansen, A.~Molinari and M.~Nardi,
  arXiv:0708.0174 [hep-ph]. K.~Itakura, O.~Morimatsu and H.~Otomo,
  Phys.\ Rev.\  D {\bf 77}, 014014 (2008).

\bibitem{mclerran} L.~P.~Csernai, J.~I.~Kapusta and L.~D.~McLerran, Phys.\ Rev.\ Lett.\  {\bf 97},
    152303 (2006).

\bibitem{rhic} R.~A.~Lacey {\it et al.}, Phys.\ Rev.\ Lett.\  {\bf 98}, 092301 (2007).

\bibitem{chiral} K.~Paech and S.~Pratt, Phys.\ Rev.\  C {\bf 74}, 014901 (2006).

\bibitem{tuchin} D.~Kharzeev and K.~Tuchin, arXiv:0705.4280 [hep-ph].

\bibitem{polyakov} A. M. Polyakov, JETP {\bf 57}, 2144 (1969).


\bibitem{karsch} F.~Karsch, D.~Kharzeev and K.~Tuchin, Phys.\ Lett.\  B {\bf 663}, 217 (2008).

\bibitem{lgts} A.~Nakamura and S.~Sakai, Phys.\ Rev.\ Lett.\  {\bf 94}, 072305 (2005).
H.~B.~Meyer, Phys.\ Rev.\  D {\bf 76}, 101701 (2007).

\bibitem{lgtb} H.~B.~Meyer, Phys.\ Rev.\ Lett.\  {\bf 100}, 162001 (2008).

\bibitem{lgtk}  K.~Huebner, F.~Karsch and C.~Pica,
  Phys.\ Rev.\  D {\bf 78},  094501 (2008).
  
  \bibitem{hh} P.~C.~Hohenberg and B.~I.~Halperin, Rev.\ Mod.\ Phys.\  {\bf 49}, 435 (1977).

\bibitem{onuki} A. Onuki, {\it Phase Transition Dynamics} (Cambridge University Press, 2002),
    p.~277.

\bibitem{ss} D.~T.~Son and M.~A.~Stephanov, Phys.\ Rev.\  D {\bf 70}, 056001 (2004).


\bibitem{hadro}
 D. Teaney, J. Lauret and E. V. Shuryak,  Phys. Rev. Lett. {\bf 86}  4783  (2001). P. Huovinen, P. F. Kolb,
U. W. Heinz, P. V. Ruuskanen and S. A. Voloshin,  Phys. Lett. B {\bf 503}, 58 (2001).  E. V. Shuryak, Nucl. Phys. A  {\bf 750}, 64 (2005).  M. Gyulassy and L. McLerran,
 Nucl. Phys. A {\bf 750}, 30 (2005). 

\bibitem{ds}
 P. Kovtun, D. T. Son and A. O. Starinets,  JHEP {\bf 0310},  064 (2003);  Phys. Rev. Lett. {\bf 94} 111601 (2005).

\bibitem{DG} P.~Danielewicz and M.~Gyulassy, Phys.\ Rev.\  D {\bf 31}, 53 (1985).



\bibitem{nambu} Y.~Nambu and G.~Jona-Lasinio, Phys.\ Rev.\  {\bf 122}, 345 (1961); Phys.\ Rev.\
    {\bf 124}, 246 (1961).

\bibitem{review} For reviews and  applications of the NJL model to hadron physics, see e.g.,
    U.~Vogl and W.~Weise, Prog.\ Part.\ Nucl.\ Phys.\  {\bf 27}, 195 (1991).  S.~P.~Klevansky, Rev.\
    Mod.\ Phys.\  {\bf 64} (1992) 649. T.~Hatsuda and T.~Kunihiro, Phys.\ Rept.\  {\bf 247}, 221
    (1994).  M.~Buballa, Phys.\ Rept.\  {\bf 407}, 205 (2005).


\bibitem{HK} A.~Hosoya and K.~Kajantie, Nucl.\ Phys.\  B {\bf 250}, 666 (1985).


\bibitem{gavin} S.~Gavin, Nucl.\ Phys.\  A {\bf 435}, 826 (1985).

\bibitem{sr} C.~Sasaki and K.~Redlich, arXiv:0806.4745 [hep-ph].




\bibitem{arnold} P.~Arnold, C.~Dogan and G.~D.~Moore, Phys.\ Rev.\  D {\bf 74}, 085021 (2006).

\bibitem{moore} G.~D.~Moore and O.~Saremi, arXiv:0805.4201 [hep-ph].





\bibitem{sfr:prd} C.~Sasaki, B.~Friman and K.~Redlich, Phys.\ Rev.\  D {\bf 75}, 054026 (2007);
    Phys. \ Rev. \ D {\bf 77}, 034024 (2008).


\bibitem{reif} F.~Reif, {\it Fundamentals of Statistical and Thermal Physics} (McGraw-Hill, New
    York, 1965).







\end{thebibliography}
\end{document}